\documentclass[showpacs,prb,preprintnumbers,amsmath,amssymb,twocolumn]{revtex4-1}
\usepackage[latin1]{inputenc}
\usepackage{graphicx}
\usepackage{dcolumn}
\usepackage{bm}
\usepackage[usenames]{xcolor}

\begin{document}

\title{Renormalization group analysis of the pair-density-wave and charge order within the fermionic hot-spot model for cuprate superconductors}
\author{Hermann Freire$^{1}$}
\email{hermann\_freire@ufg.br}
\author{Vanuildo S. de Carvalho$^{1,2}$}
\author{Catherine Pépin$^{2}$}
\affiliation{$^{1}$ Instituto de Física, Universidade Federal de Goiás, 74.001-970, Goiânia-GO, Brazil}
\affiliation{$^{2}$ IPhT, L'Orme des Merisiers, CEA-Saclay, 91191 Gif-sur-Yvette, France}
\date{\today}

\begin{abstract}
In light of the new experimental and theoretical important developments in high-$T_c$ superconductivity,
we revisit the fermionic hot-spot model relevant to the phenomenology of the
cuprates. We extend previous results by means of a complete two-loop order renormalization group
(RG) framework. Here, we explicitly study the effect of the charge-density-wave (CDW) order
parameter with a $d$-wave form factor with the experimentally observed modulation $(\pm Q_0,0)$ and $(0,\pm Q_0)$ 
at the infrared-stable nontrivial fixed point obtained previously for this model. Additionally, we proceed to investigate 
also the so-called pair-density-wave (PDW) order that was recently proposed in the literature as
a possible candidate for the ``hidden'' order to describe the pseudogap phase observed in underdoped cuprates.
We confirm that although the above two ordering tendencies are also found to be nearly degenerate both at one-loop and two-loop RG orders
and linked by an emergent $SU(2)$ pseudospin symmetry,
they turn out to be subleading for weaker couplings in the present model to antiferromagnetism, $d$-wave
bond-density wave (BDW) order with modulation along Brillouin zone diagonals $(\pm Q_0,\pm Q_0)$,
and $d$-wave singlet superconductivity (SSC). However, as we increase the strength of the initial coupling towards moderate values, we do capture
a tendency for the entangled PDW/CDW order to become leading compared to BDW/SSC in the model,
which suggests that the former composite order might be indeed a viable concept to describe  
some cuprate superconductors at high temperatures in the underdoped regime, as has been recently alluded to by many authors in the literature. 
\end{abstract}


\maketitle

\section{Introduction}

The underlying nature of the pseudogap phase observed in the underdoped cuprates continues to be one
of the most enigmatic and profound problems in condensed matter theory. Despite this statement, a lot has been learned about this phase
both theoretically and experimentally in the last years.
On the experimental side, groundbreaking measurements performed on the non-Lanthanum-based materials 
have increased dramatically our present knowledge about these compounds. Nuclear magnetic resonance experiments \cite{Julien,Julien2}, 
resonant \cite{Ghiringhelli,Achkar} and hard \cite{Chang2} x-ray scattering, ultrasound 
measurements \cite{LeBoeuf} and scanning tunneling microscopy \cite{Hoffman,Yazdani} confirmed the emergence of a charge-density-wave order (CDW)
at small hole doping at wavevectors $(\pm Q_0,0)$ and $(0,\pm Q_0)$ oriented along the principal axes of the CuO$_2$ planes with 
a predominant $d$-wave form factor \cite{Comin,Fujita}. This CDW has
the form of a checkerboard and is incommensurate with the lattice, since it connects approximately the ``tips'' of the Fermi arcs \cite{Comin2,SilvaNeto}.
This gave an initial impetus for the interpretation that such a CDW could be the missing ``hidden'' order in the pseudogap phase that might be ultimately
responsible for the reconstruction of the Fermi surface observed, e.g., in quantum oscillations experiments in YBCO
at high magnetic fields \cite{Taillefer,Sebastian}. However, this latter interpretation, at least at face value,
has been recently challenged by some experiments due to the fact that
the transition line of this CDW phase is essentially dome-shaped and, in general, is not coincident with the pseudogap line $T^*$ observed
in the phase diagram of these compounds. 

Since the ``tips'' of the Fermi arcs in the pseudogap phase
are close to the so-called hot spots (i.e., the points in momentum space in which 
the antiferromagnetic zone boundary intersects a putative underlying Fermi surface of these compounds),
this could mean that a possible low-energy effective model to describe these materials might be a spin-fermion model, or more generally, a hot-spot model \cite{Chubukov,Sachdev,Efetov,LaPlaca,Efetov_2,Chowdhury,Sau}.
In this respect, it has been famously demonstrated in Ref. \cite{Sachdev} that, if the energy dispersion is linearized, an emergent $SU(2)$ pseudospin symmetry that rotates the
$d$-wave superconducting order parameter onto a ``$d$-wave bond-density wave'' (BDW) order\cite{LaPlaca} with modulation along Brillouin zone diagonals $(\pm Q_0,\pm Q_0)$
exists in the model and this property effectively produces a composite order parameter with both bond order and preformed pairs at high temperatures \cite{Efetov}.
However, a generally acknowledged drawback of this approach is related to the fact that the leading charge order
indeed always appears at the wavevector $(\pm Q_0,\pm Q_0)$, which, so far, has never been observed experimentally.
Many alternative scenarios \cite{Wang,Allais,Allais_2,Tsvelik,Kampf,Chowdhury_2,Thomson,Bulut} that aim to resolve this discrepancy with experiments have also been proposed in the literature,
but no consensus has yet emerged in the community with regard to what is the mechanism responsible
for the pseudogap phase in the cuprates. 

For this reason, many researchers continue trying to uncover what might be the ``hidden'' order that is fundamentally responsible 
for the emergence of the pseudogap phase seen in the cuprates. Several good candidates on the table (which are not mutually exclusive) 
include pair-density-wave (PDW) order \cite{PALee,Agterberg,Fradkin,Wang_2},
the aforementioned $d$-wave CDW order at wavevectors $(\pm Q_0,0)$ and $(0,\pm Q_0)$ (which may also break additional discrete symmetries such as the $C_4$ lattice rotational symmetry down to the
$C_2$, and time-reversal and parity symmetries \cite{Wang,Tsvelik}), Mottness  \cite{PALee,PALee_2}, orbital loop current order \cite{Bourges,Bourges_2,Varma}, fractionalized Fermi liquid \cite{Chowdhury_2}, among others \cite{Bulut}.
For simplicity, we shall concentrate in this work only on the first two candidate orders (i.e., PDW and CDW) and leave the analysis of other
possible orders that might be present in the underdoped cuprates for a future study.  

The PDW order refers essentially to a superconducting order with a finite Cooper-pair 
center of mass momentum, similar in this respect to a Fulde-Ferrell-Larkin-Ovchinnikov (FFLO) state \cite{Fulde,Larkin}.
It has been recently proposed that this order may account for the anomalous quasiparticle excitations seen by angle-resolved photoemission (ARPES)
experiments in both LBCO and YBCO\cite{Fradkin}. The PDW order could also potentially emerge from a doped Mott insulator scenario
as a strong-coupling instability due to an effect similar to ``Ampere's law" \cite{PALee}.
In this respect, in  Ref. \cite{PALee} it was argued that the pseudogap phase in underdoped cuprates is more appropriately described by a PDW instead of simply a CDW.
Additionally, it has also been pointed out in the literature that the PDW order might give rise to a secondary 
non-superconducting order parameter 
that breaks both time-reversal and parity symmetries, but preserves their product \cite{Agterberg}.
The breaking of time-reversal and parity symmetries in the pseudogap regime would be consistent, e.g., 
with the experimental signatures obtained both via polarized neutron scattering \cite{Greven} and Kerr-rotation experiments \cite{Kapitulnik,Kapitulnik_2}.
Therefore, it becomes clearly important to analyze, in a more detailed way, the role of the PDW order parameter in the context of low-energy effective models that
might describe the essence of the physics of the underdoped cuprates.

\begin{figure}[t]
 \includegraphics[height=2.7in]{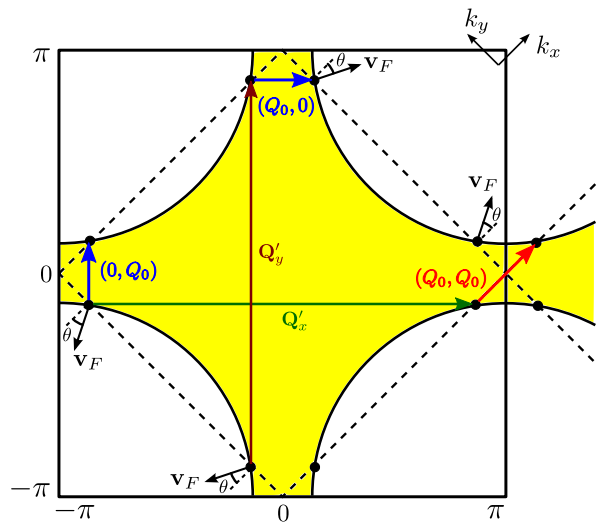}
 \caption{(Color online) The present 2D fermionic model consisting of eight hot spots (with bare Fermi velocities $\mathbf{v_F}$
 and angle $\theta$) on the Fermi surface
 that are directly connected by the commensurate spin density wave (SDW) ordering wavevector $\mathbf{Q}=(\pi,\pi)$. 
 The incommensurate wavevectors $\mathbf{Q}_{x}=(Q_0,0)$, $\mathbf{Q}_{y}=(0,Q_0)$, $\widetilde{\mathbf{Q}}=(Q_0,Q_0)$,
 $\mathbf{Q}^{'}_{x}$ and $\mathbf{Q}^{'}_{y}$ are also displayed.}
 \label{fig:Fermi_surface}
\end{figure}

In light of the many recent experimental and theoretical developments described above, 
we build upon a work by two of us by revisiting here the so-called fermionic hot spot model defined in Ref. \cite{Freire5} 
and analyzed within a two-loop renormalization group (RG)
framework. The fermionic hot-spot model can be seen as descendant of the Abanov-Chubukov spin-fermion model\cite{Chubukov} that, most importantly, includes
here all relevant interactions between the fermions and, for this reason, enables one to analyze all possible instabilities of the 
model from weak to moderate couplings on equal footing.
We mention that the fermionic hot spot model has
also been recently studied by Whitsitt and Sachdev \cite{Seth} within a one-loop RG approach and their results, up to this order, agree with ours.
In the present paper, we explicitly study the effect of the charge-density-wave (CDW) order
parameter with a $d$-wave form factor with the experimentally observed wavevectors $(\pm Q_0,0)$ and $(0,\pm Q_0)$ 
in the vicinity of the infrared-stable nontrivial fixed point obtained at two loops \cite{Freire5} that, as a result, 
naturally implies a new quantum critical universality 
class for the present model. Moreover, we proceed to investigate in this scenario
also the role of the PDW order, which was recently proposed in the literature as
a viable candidate for the ``hidden'' order to describe the pseudogap phase observed in the underdoped cuprates.
Interestingly, we confirm that the above two orders turn out to be also nearly degenerate both at one-loop and two-loop RG orders
and clearly linked by an emergent $SU(2)$ pseudospin symmetry at low energies. Despite this fact,
they continue to be subleading for weaker couplings in the present model to antiferromagnetism, $d$-wave
bond-density wave (BDW) order with modulation given by $(\pm Q_0,\pm Q_0)$
and $d$-wave singlet superconductivity (SSC). We then finish by discussing how the PDW/CDW order is likely to become
leading compared to BDW/SSC order in this model, which would potentially agree with the experimental situation.

This paper is organized as follows. In Section II, we define the so-called fermionic hot spot model that, as we will explain,
might be relevant to describe some aspects of the phenomenology of the underdoped cuprates. In Section III,
we briefly introduce the field-theoretical RG approach and show how to implement this method up to two loops in
the present model. In this part, we will concentrate specifically on discussing the role played
by the PDW and CDW orders (both with a predominant $d$-wave factor) at the experimentally observed wavevectors $(\pm Q_0,0)$ and $(0,\pm Q_0)$ 
that emerge in the low-energy limit of the model.
The relevant RG flow equations are then derived and solved in Section IV. Finally, we present our conclusions in Section V.

\begin{figure*}[t]
 \includegraphics[height=2.8in]{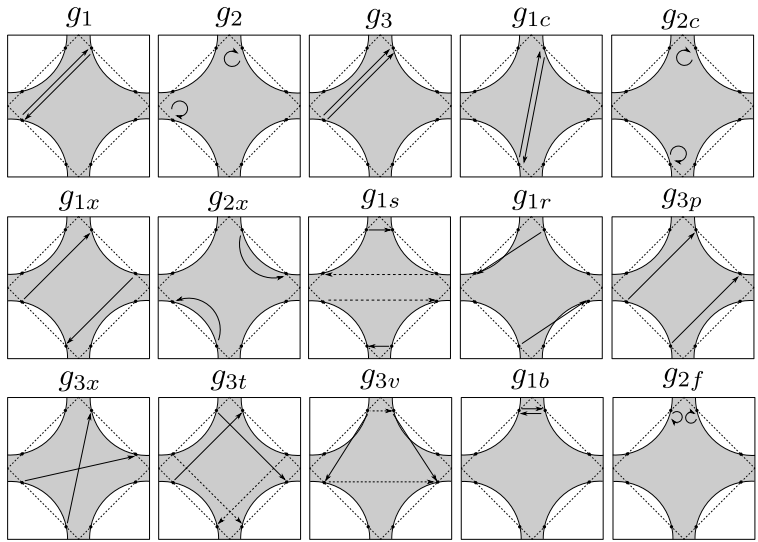}
 \caption{The relevant couplings in the present fermionic hot-spot model.
For the $g_{1s}$, $g_{3t}$ and $g_{3v}$ couplings, we show two possible scattering
processes (one by a solid arrow and the other by a dashed arrow),
which are always equal in our two-loop RG results.}
 \label{fig:Interactions}
\end{figure*}

\section{The Fermionic Hot-Spot Model}

To begin with, we consider a single-band fermionic model whose noninteracting part is given by the following energy dispersion 
$\xi_{\mathbf{k}}=-2t(\cos(k_x)+\cos(k_y))-4t'\cos(k_x)\cos(k_y)-\mu$, which is given in terms of units of the lattice constant, 
the parameters $t$ and $t'$ are, respectively, the nearest neighbor and next-nearest neighbor hoppings and $\mu$ is the
chemical potential. For the specific case of the cuprates, one may assume for simplicity the choice of parameter $t'=-0.3t$, which, at low hole doping, 
results in the Fermi surface (FS) displayed in Fig. 1. 
In the interval $0<|\mu|<4|t'|$, the corresponding FS intersects the antiferromagnetic zone boundary at eight points (the aforementioned hot spots).
These hot spots are shown in Fig. 1
and the excitations residing near the locus of these points in fact generate the most singular
contributions in the model at long distances and long timescales. Besides, one can naturally rotate the momentum axes  
by an angle of $\pi/4$
by defining new axes $({k}_{x},{k}_{y})$, where the rotated momenta ${k}_{x}$ and ${k}_{y}$ are depicted in Fig. 1.

Since we will focus on the universal properties of this
model, we may linearize the energy dispersion around the hot spots as $\xi_{\mathbf{k}}\approx \mathbf{v_F}.\mathbf{k}$
with $\mathbf{k}$ being the momentum measured relative to the hot spots, 
the Fermi velocity $\mathbf{v_F}$=($v_{x}$,$v_{y}$) with the modulus of the components given by 
$v_{x}=|\nabla_{(k_{x},k_{y})}\xi_{\mathbf{k}}|_{\mathbf{k}=\mathbf{k}_F}|
\cos{\theta}$ and $v_{y}=|\nabla_{(k_{x},k_{y})}\xi_{\mathbf{k}}|_{\mathbf{k}=\mathbf{k}_F}|
\sin{\theta}$, with $\theta$ being the angle depicted in Fig. 1. Here, we specialize to the value $\mu\approx -0.77t$
(i.e., a low hole-doped regime). 
It is interesting to point out, however, that our results are fairly
insensitive to this choice of the chemical potential, as long as $0<|\mu|<4|t'|$.
The components of the momentum $k_x$ and $k_y$ are restricted to the interval $[-k_{c},k_{c}]$, where $k_{c}$
stands for a sharp ultraviolet (UV) momentum cutoff in our theory. 
This implies also a high-energy cutoff, which is given by $\Lambda_0=2v_{F}^{R}k_{c}$ that
we choose to be of the same order of the full bandwidth of the problem, i.e., $\Lambda_0\sim 8t$.

The fermionic hot spot model at zero temperature is described by the 
partition function $\mathcal{Z}=\int \mathcal{D}[\overline{\psi},\psi] \exp({i\int_{-\infty}^{\infty} dt\, L_{R}[\overline{\psi},\psi]})$ with the
interacting renormalized Lagrangian $L_{R}$ given by

\vspace{-0.2cm}

\begin{align}\label{lagrangian}
L_{R}&=\sum_{\mathbf{k},\sigma}Z_{\psi}\,\overline{\psi}^{R}_{\sigma}(\mathbf{k})\big[i\partial_t-Z_{v}^{x}v_{x}^{R}k_x-Z_{v}^{y}v_{y}^{R}k_y
\big]\psi^{R}_{\sigma}(\mathbf{k})\nonumber\\
&-\sum_{i}\sum_{\substack{{\mathbf{k_1,k_2,k_3}} \\ {\sigma,\sigma'}}}
Z^{2}_{\psi}\,g_{i}^{B}\,\overline{\psi}_{\sigma}^{R}(\mathbf{k_4})\overline{\psi}_{\sigma'}^{R}(\mathbf{k_3})\psi_{\sigma'}^{R}(\mathbf{k_2})\psi_{\sigma}^{R}(\mathbf{k_1}),
\end{align}

\noindent where $\mathbf{k_4}=\mathbf{k_1}+\mathbf{k_2}-\mathbf{k_3}$ and the volume $V$ has been set equal to unity. 
All bare quantities
(indicated by the superscript $B$) are related to the renormalized quantities (indicated by the superscript $R$) by the following standard expressions: 
$\psi_{\sigma}^{B}(\mathbf{k})= Z^{1/2}_{\psi}\psi_{\sigma}^{R}(\mathbf{k})$, 
$\overline{\psi}_{\sigma}^{B}(\mathbf{k})= Z^{1/2}_{\psi}\overline{\psi}_{\sigma}^{R}(\mathbf{k})$, 
$v_{x}^{B}=Z_{v}^{x}v_{x}^{R}$ and $v_{y}^{B}=
Z_{v}^{y}v_{y}^{R}$, where $Z_{\psi}$ is the 
field renormalization factor and $Z_{v}^{x}$ and $Z_{v}^{y}$ are the Fermi velocity renormalization factors. The renormalized Grassmann fields
$\overline{\psi}_{\sigma}^{R}(\mathbf{k})$ and $\psi_{\sigma}^{R}(\mathbf{k})$ are associated, respectively, to the creation and annihilation operators
of the low-energy excitations residing in the vicinity of the hot spots with momentum $\mathbf{k}$ and spin projection $\sigma$. The index $i$ refers to 
the many possible interaction processes in
the model that generate singularities within perturbation theory. Here we follow roughly the notation of Ref. \cite{Freire5}, i.e., 
$i=1,2,3,1c,2c,1x,2x,1s,1r,3x,3p,3t,3v,1b,2f$ (for details
of the couplings taken into account, see Fig. 2). In this way, our approach will mirror
the so-called ``g-ology'' notation \cite{Solyom}, adapted to our 2D problem at hand.
Additionally, we define dimensionless renormalized couplings (denoted by $\bar{g}_{i}^R$)
in the following way: $g_{i}^{B}=N_{0}^{-1}Z^{-2}_{\psi} \left[\bar{g}_{i}^{R}+\delta\bar{g}_{i}^{R}\right]$, where $N_0=k_c/\pi^2 v_{F}^{R}$
is the density of states at the Fermi level. 
In all above expressions, we set 
conventionally $Z_{\psi}=1+\delta Z_{\psi}$, $Z_{v}^{x}=Z_{\psi}^{-1}(1+\delta Z_{v}^{x})$, 
$Z_{v}^{y}=Z_{\psi}^{-1}(1+\delta Z_{v}^{y})$, where $\delta Z_{\psi}$, $\delta Z_{v}^{x}$,
$\delta Z_{v}^{y}$ and $\delta \bar{g}_{i}^{R}$ are
the counterterms that will be calculated order by order within the  
renormalized perturbation theory \cite{Peskin}.

Since the number of couplings in the present model is large, the possibilities for choosing their initial conditions
are also considerable. Therefore, our choice here will be motivated by
relevant physical microscopic models. We shall restrict our analysis in the present paper to the paradigmatic 2D Hubbard model, whose initial conditions 
for the couplings are given by $g_{i}^R=k_c U/(\pi^2 v_F^R)$ for all $i$, where $U>0$ is
the local on-site repulsive interaction strength.

\section{Field-theoretical RG}

In this section, we describe the two-loop RG approach that we shall apply in the rest of this work.
Our discussion here will be relatively concise, 
since the field-theoretical RG methodology was already explained by two of us in great detail 
in the context of the present model elsewhere \cite{Freire5,Freire2}.
Within perturbation theory, if we compute many one-loop quantities, such as, e.g., particle-particle and particle-hole 
polarization bubbles for several
choices of incoming external momenta $\mathbf{q}$ and also their corresponding two-loop-order corrections,
we obtain several logarithmic divergences of the type $\ln(\Lambda_0/\Lambda)$
as we probe the system towards the low-energy limit $\Lambda\rightarrow 0$ (see also Refs. \cite{Ferraz,Freire,Freire4}).
As we have already explained above, we circumvent this problem by defining counterterms 
for all the dimensionless couplings, the fermionic fields and the Fermi velocity of the model that, by construction, effectively regulate all divergences order by order within 
the renormalized perturbation theory. The resulting coupled-differential RG equations obtained by this method are thus given in Appendix A.

To exemplify the field-theoretical RG method, if we calculate the self-energy of the present model (e.g., for $k_x >0$ and $k_y >0$) at two-loop order within the renormalized perturbation theory, 
we obtain the following nonanalytic contribution

\vspace{-0.4cm}

\begin{align}
&\Sigma_{R}^{(\text{2 loops})}(\mathbf{k},\omega)=-\frac{k_c^2}{4\pi^2(v_F^{R})^{2}}\left[\gamma_{\psi}\omega-\gamma_{v_x}v_x^R k_x
+\gamma_{v_y}v_y^R k_y\right]\nonumber\\
&\times \ln\left(\frac{\Lambda_0}{\text{max}\{\omega,v_F^R |\mathbf{k}|\}}\right)-\left[\delta Z_{\psi}\,\omega-\delta Z_{v}^{x}v_x^R k_x
-\delta Z_{v}^{y}v_y^R k_y\right],
\end{align}

\noindent where $\gamma_{\psi}=\gamma_{v_x}=(g_{1c}^2+g_{2c}^2+g_{1x}^2+g_{2x}^2+g_{3p}^2+g_{3x}^2-g_{1c}g_{2c}-g_{1x}g_{2x}-g_{3p}g_{3x})$
and $\gamma_{v_y}=(g_{1x}^2+g_{2x}^2+g_{3p}^2+g_{3x}^2-g_{1c}^2-g_{2c}^2+g_{1c}g_{2c}-g_{1x}g_{2x}-g_{3p}g_{3x})$.
For simplicity, we shall omit the superscript $R$ in the above renormalized dimensionless couplings of the model from here on. As a result,
we obtain the following RG flow equation up to two loops associated with the renormalized Fermi velocity $\mathbf{v}_F^R=(v_x^R,v_y^R)$, i.e.,

\vspace{-0.3cm}

\begin{align}\label{RG_vF}
\Lambda \frac{d\,\kappa_R}{d\Lambda}=\frac{\kappa_R}{4(1+\kappa_R^2)}\widetilde{\gamma}_{v_y}(\{g_i^R\}),
\end{align}

\noindent where $\kappa_R=(v_y^R/v_x^R)=\tan\theta_R$ is the ratio between the components of the Fermi velocity at the hot spots
and $\widetilde{\gamma}_{v_y}(\{g_i^R\})=\gamma_{v_y}+\gamma_{\psi}$. From Eq. (\ref{RG_vF}), one obtains
that the only infrared (IR) stable fixed point corresponds to $\kappa_R^*\rightarrow 0$ (or $\theta_R^*\rightarrow 0$), which refers to
a perfect nesting condition of the hot spots connected by the antiferromagnetic (AF) ordering wavevector $\mathbf{Q}=(\pi,\pi)$.
Although this only happens, strictly speaking, in the limit of $\Lambda\rightarrow 0$, one can check numerically that for
moderate couplings and not too low 
RG scales, $\kappa$ is already reasonably suppressed and the resulting renormalized FS exhibits an approximately nesting
condition connected by $(\pi,\pi)$ (see, e.g., Fig. 3). This fact will be clearly favorable to antiferromagnetic fluctuations
that will play a major role
in the present system as will become clear shortly.

It is important to discuss at this point on the
contribution of both particle-particle and particle-hole bubbles
for some choices of wavevector $\mathbf{q}$ (to check our present definitions, see Fig. 1).
If we calculate those quantities, we obtain the following expressions


\vspace{-0.3cm}

\begin{align}
\Pi_{PH}(\mathbf{q}=\mathbf{\widetilde{Q}})&=\frac{N_0 }{2}\ln\left(\frac{\Lambda_0}{\Lambda}\right),\label{PH1}\\
\Pi_{PP}(\mathbf{q}=\mathbf{\widetilde{Q}})&=-\frac{N_0 }{2}\ln\left(\frac{\Lambda_0}{\Lambda}\right),\label{PP1}\\
\Pi_{PH}(\mathbf{q}=\mathbf{Q})&\approx\frac{N_0 }{2 \cos\theta_R}\ln\left(\frac{\Lambda_0}{\text{max}\{\Lambda,2v_F^R k_c\sin\theta_R\}}\right),\\
\Pi_{PH}(\mathbf{q}=\mathbf{Q}_x)&\approx N_0 \lambda\ln\left(\frac{\Lambda_0}{\text{max}\{\Lambda,\frac{v_F^R k_c(1-\tan\theta_R)}{\lambda}\}}\right),\label{PH}\\
\Pi_{PP}(\mathbf{q}=\mathbf{Q}_x)&\approx -N_0 \lambda\ln\left(\frac{\Lambda_0}{\text{max}\{\Lambda,\frac{v_F^R k_c(1-\tan\theta_R)}{\lambda}\}}\right),\label{PP}\\
\Pi_{PH}(\mathbf{q}=\mathbf{Q}'_x)&\approx N_0 \lambda'\ln\left(\frac{\Lambda_0}{\text{max}\{\Lambda,\frac{v_F^R k_c(1+\tan\theta_R)}{\lambda'}\}}\right),\\
\Pi_{PP}(\mathbf{q}=\mathbf{Q}'_x)&\approx -N_0 \lambda'\ln\left(\frac{\Lambda_0}{\text{max}\{\Lambda,\frac{v_F^R k_c(1+\tan\theta_R)}{\lambda'}\}}\right),
\end{align}


\noindent with $\Lambda$ being the RG scale which could be interpreted (to logarithmic accuracy) in terms of an effective temperature scale. 
Besides, the prefactors $\lambda$ and $\lambda'$ are given by $\lambda=1/(\cos\theta_R +\sin\theta_R)$ and $\lambda'=1/(\cos\theta_R -\sin\theta_R)$ for the present hot spot model. We mention here that even though the bubbles $\Pi_{PH}(\mathbf{q}=\mathbf{Q}_x)$ and $\Pi_{PP}(\mathbf{q}=\mathbf{Q}_x)$ have initially different prefactors and different low-energy cutoffs compared to $\Pi_{PH}(\mathbf{q}=\mathbf{Q}'_x)$ and $\Pi_{PP}(\mathbf{q}=\mathbf{Q}'_x)$, those prefactors and cutoffs become identical
for the renormalized Fermi surface (i.e., $\theta_R^*\rightarrow 0$). In this way,
we will refer only to one type of particle-particle and particle-hole bubbles [i.e., $\Pi_{PH}(\mathbf{q}=\mathbf{Q}_x)$ and $\Pi_{PP}(\mathbf{q}=\mathbf{Q}_x)$] in what follows. All the above polarization functions agree precisely with those derived in Ref. \cite{Seth}.
Those particle-hole and particle-particle bubbles at moderate energy scales 
satisfying $\Lambda>2v_F^R k_c\sin\theta_R$ and $\Lambda>v_F^R k_c(1-\tan\theta_R)/\lambda$
(which would correspond physically to a ``high-temperature'' regime)
will approximately generate logarithmic divergences as a function of $\Lambda$ that will allow us to use the field-theoretical RG approach to this model.  
A similar result also holds for 
the particle-particle and particle-hole bubbles at $\mathbf{Q}_y=(0,Q_0)$ and $\mathbf{Q}'_y$. 
For moderate interactions,
it can be shown numerically that a nontrivial fixed point can be reached within such a high-energy regime.
We emphasize here that such an approximation
resonates with the idea that there is possibly a
`high-temperature' pseudogap ($T<T^*$) in the cuprate superconductors, which
plays the role of the `normal' state in these compounds out of which many broken symmetry states
(such as, e.g., a phase with either incommensurate charge order or possibly also pair-density-wave order) emerge
at lower temperatures ($T<T^{**}< T^*$).

\begin{figure}[t]
 \includegraphics[height=2.1in]{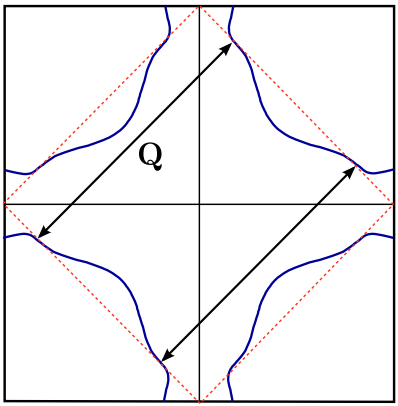}
 \caption{(Color online) A possible renormalized Fermi surface suggested by the present two-loop RG calculation
 of the fermionic hot-spot model. Note the resulting nesting condition of the hot spots connected by $\mathbf{Q}=(\pi,\pi)$.}
 \label{fig:Fermi_surface2}
\end{figure} 

We first begin by reviewing previous results\cite{Freire5} and then we present here new results concerning the present model. 
We integrate numerically the RG flow equations for the model [Eqs. (A1)--(A15)] using conventional fourth-order Runge-Kutta method.
In one-loop RG order, the numerical solution of those equations shows that, 
despite the fact that there is no evidence of a non-trivial fixed point in the model at this order, nearly all dimensionless coupling 
constants diverge at the same critical RG step value $l_c = \ln(\Lambda_0/\Lambda_c)$, which depends only on the initial conditions 
for the couplings. This divergence is appropriately described by the following scaling ansatz $g_i (l)=C_i/(l_c-l)$, with $l_c\propto (1/U)$ and
the $C_i$ are universal constants that do not depend sensitively on the values of the
initial conditions. For instance, for Hubbard-like initial conditions, their numerical values are found to be

\vspace{-0.3cm}

\begin{align}
&(C_{1},C_{2},C_{3},C_{1c},C_{2c},C_{1x},C_{2x}, 
C_{3x},C_{3p},C_{3t},C_{3v})\approx \nonumber\\
&(0.07,0.23,0.39,-0.04,-0.02,0.13,0.19,0.19,0.11,\nonumber\\
&0.12,0.25),
\end{align}

\noindent and $C_{1s}=C_{1r}=C_{1b}=C_{2f}=0$. It is interesting to note that in the resulting effective theory there are indeed only 
four coupling constants that do not diverge ($g_{1s}$, $g_{1r}$, $g_{1b}$,
and $g_{2f}$) as the RG scale $\Lambda$ is lowered towards $\Lambda_c$. In this way, they become comparatively much smaller than the other couplings
at low energies. For this reason, one can go back to the original model [Eq. (\ref{lagrangian})] 
and neglect for simplicity all these irrelevant couplings to begin with.
We shall perform this in the remainder of this work.

Next, we move on to the two-loop RG case. An important point we wish to stress here is
that it is crucial to implement the RG method for the present model at least at two loops or beyond, since as we have 
shown previously the first nonanalytic contribution to the self-energy (and, consequently, the renormalized single-particle
Green's function and related quantities\cite{Freire5,Ferraz}) only emerges at this order. 
In this respect, by solving numerically the complete two-loop RG equations [Eqs. (A5)-(A15)], we can show that one finds instead
a ``landscape'' of many nontrivial fixed points in the model, in addition of course to the trivial (i.e., Fermi-liquid) one.
Interestingly, only one nontrivial fixed point \cite{Freire5} turns out to be stable in all directions in the IR regime, so that 
any trajectory in the coupling parameter space close to this fixed point will necessarily converge to it as the RG scale $\Lambda\rightarrow 0$.
For this reason, we shall focus our attention from here on only on this single IR-stable nontrivial fixed point,
since it naturally implies a new quantum critical universality class for the present problem. The numerical results for
this nontrivial fixed point obtained at two-loop order are approximately given by

\vspace{-0.3cm}

\begin{align}
&(g^{*}_{1},g^{*}_{2},g^{*}_{3},g^{*}_{1c},g^{*}_{2c}, 
g^{*}_{1x},g^{*}_{2x},g^{*}_{3x},g^{*}_{3p},g^{*}_{3t},g^{*}_{3v})\approx\nonumber\\
&(0,1.68,1.84,-2.0,-1.0,1.92,1.92,1.92,0,0,0).
\end{align}

\noindent We will refer to the above nontrivial fixed point henceforth as simply the hot-spot fixed point (HSFP). It is important to emphasize that Umklapp interactions in the present fermionic hot-spot model are fundamental for stabilizing the HSFP in the IR-limit.
It may also be noteworthy the fact that some couplings, due to two-loop-order quantum fluctuations, now flow asymptotically to zero
at low energies, whereas others flow instead to finite values. 
This interesting fact should potentially simplify the solution of the present model by means of other 
complementary methods.

\section{Linear Response Theory}

To study the enhanced correlations and possible ordering tendencies in the present model,
we must add to Eq. (1) the following term

\vspace{-0.3cm}

\begin{align}\label{external}
&L_{ext}=\sum_{\mathbf{k},\alpha,\beta}\Delta_{B,SC}^{\alpha\beta}(\mathbf{k},\mathbf{q})\,\overline{\psi}_{B\alpha}(\mathbf{k}+\mathbf{q}/2)\overline{\psi}_{B\beta}(-\mathbf{k}+\mathbf{q}/2) \nonumber\\
&+\sum_{\mathbf{k},\alpha,\beta}\Delta_{B,DW}^{\alpha\beta}(\mathbf{k},\mathbf{q})\,\overline{\psi}_{B\alpha}(\mathbf{k}+\mathbf{q}/2)\psi_{B\beta}(\mathbf{k}-\mathbf{q}/2)
+ H.c.,
\end{align}

\noindent where $\Delta_{B,SC}^{\alpha\beta}(\mathbf{k},\mathbf{q})$ and $\Delta_{B,DW}^{\alpha\beta}(\mathbf{k},\mathbf{q})$ correspond to the bare response vertices
for the superconducting (SC) and density-wave (DW) orders, respectively. This
additional term will generate some new Feynman diagrams with three-legged vertices (see Fig. 4),
which may also exhibit logarithmic divergences as a function of the RG scale $\Lambda$ in the model.
We can then conventionally define the renormalized
response vertices with their corresponding counterterms
as follows: 
$\Delta_{B,i}^{\alpha\beta}(\mathbf{k},\mathbf{q})=Z^{-1}_{\psi}
[\Delta_{R,i}^{\alpha\beta}(\mathbf{k},\mathbf{q})+\delta \Delta_{R,i}^{\alpha\beta}(\mathbf{k},\mathbf{q})]$ for $i=SC$ and $DW$. 
As a result, we shall obtain straightforwardly the corresponding RG equations up to two loops. Since two of us have already derived elsewhere \cite{Freire5}
the RG equations for the order parameters associated with antiferromagnetism, $d$-wave superconductivity and $d$-wave bond-density-wave order
at the incommensurate wavevector $\mathbf{\widetilde{Q}}=(Q_0,Q_0)$, we shall not repeat them here.
For this reason, we concentrate 
only on the PDW and $d$-wave charge order response vertices with modulation $\mathbf{Q}_x=(Q_0,0)$ [or $\mathbf{Q}_y=(0,Q_0)$]
in what follows. If we take into account the effect of the renormalization
of the FS, the corresponding RG equations are given by

\vspace{-0.1cm}

\begin{align}\label{DW2}
&\Lambda\frac{d\Delta_{R,DW(\mathbf{Q}_{x(y)})}^{(1)\alpha\beta}}{d\Lambda}=\lambda\bigg[
g_{3v}\sum_{\sigma=\alpha,\beta}\Delta_{R,DW(\mathbf{Q}_{x(y)})}^{(2)\sigma\sigma}\nonumber\\
&-g_{3t}\Delta_{R,DW(\mathbf{Q}_{x(y)})}^{(2)\beta\alpha}\bigg]+\eta\,\Delta_{R,DW(\mathbf{Q}_{x(y)})}^{(1)\alpha\beta},
\end{align}

\begin{align}\label{PDW} 
\Lambda\frac{d\Delta_{R,SC(\mathbf{Q}_{x(y)})}^{(1)\alpha\beta}}{d\Lambda}&=
\lambda\left[g_{3t}\Delta_{R,SC(\mathbf{Q}_{x(y)})}^{(2)\alpha\beta}-g_{3v}\Delta_{R,SC(\mathbf{Q}_{x(y)})}^{(2)\beta\alpha}\right]\nonumber\\
&+\eta\,\Delta_{R,SC(\mathbf{Q}_{x(y)})}^{(1)\alpha\beta},
\end{align}

\begin{figure}[t]
 \includegraphics[height=1.8in]{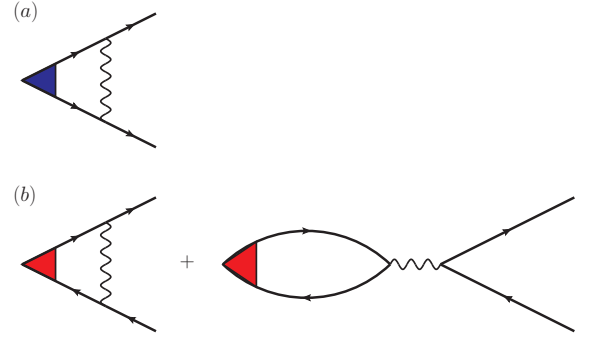}
 \caption{(Color online) Feynman diagrams for (a) SC and (b) DW response
functions of the present fermionic hot-spot model. 
Solid lines represent noninteracting fermionic single-particle Green's function, while the wavy lines stand for the renormalized coupling constants. 
Depending on the channel analyzed, the vector $\mathbf{q}$ can be: $\mathbf{Q}=(\pi,\pi)$, 
$\mathbf{\widetilde{Q}}=(Q_0,Q_0)$,  $\mathbf{Q}_x=(Q_0,0)$, $\mathbf{Q}_y=(0,Q_0)$, $\mathbf{Q}'_x$ and $\mathbf{Q}'_y$.}
 \label{fig:response_function}
\end{figure}

\noindent where the two-loop self-energy feedback contribution is given by the anomalous dimension 
$\eta=\Lambda (d\ln Z_{\psi}/d\Lambda)=(g^{2}_{1}+g^{2}_{2}+g^{2}_{1c}+g^{2}_{2c}+g^{2}_{1x}+g^{2}_{2x}
-g_{1}g_{2}-g_{1c}g_{2c}-g_{1x}g_{2x}-g_{3p}g_{3x}+g^{2}_{3p}
+g^{2}_{3x}+{g^{2}_{3}}/{2})/4$. It is interesting to point out here that the above RG equations are
the same for both $\mathbf{Q}_{x}$ and $\mathbf{Q}_{y}$, such that the $C_4$ lattice rotational symmetry is always preserved in our theory.
This explains our present notation $\mathbf{Q}_{x(y)}$ in the response vertices. 

By antisymmetrizing these response vertices with respect to the spin indices, we obtain the PDW and CDW order parameters

\begin{center}
$\left\{%
\begin{array}{ll}
    \Delta^{(j)}_{PDW(\mathbf{Q}_{x(y)})}=\Delta_{R,SC(\mathbf{Q}_{x(y)})}^{(j)\uparrow\downarrow}-\Delta_{R,SC(\mathbf{Q}_{x(y)})}^{(j)\downarrow\uparrow}\\
    {\Delta}^{(j)}_{CDW(\mathbf{Q}_{x(y)}))}={\Delta}_{R,DW(\mathbf{Q}_{x(y)})}^{(j)\uparrow\uparrow}+{\Delta}_{R,DW(\mathbf{Q}_{x(y)})}^{(j)\downarrow\downarrow},
    \end{array}%
\right.$ 
\end{center}

\noindent where the points at the FS are denoted by $j=1,2$ (see Fig. 5). To determine the symmetry of
the order parameter, we must further symmetrize (or antisymmetrize) the response vertices with respect to the index $j$. Thus, we obtain

\begin{center}
$\left\{%
\begin{array}{ll}
    \Delta_{PDW(\mathbf{Q}_{x(y)})}=\Delta_{PDW(\mathbf{Q}_{x(y)})}^{(1)}-\Delta_{PDW(\mathbf{Q}_{x(y)})}^{(2)}.\vspace{0.1cm} \\
    {\Delta}_{CDW(\mathbf{Q}_{x(y)})}^{(d-wave)}={\Delta}_{CDW(\mathbf{Q}_{x(y)})}^{(1)}-{\Delta}_{CDW(\mathbf{Q}_{x(y)})}^{(2)},\vspace{0.1cm} \\
    {\Delta}_{CDW(\mathbf{Q}_{x(y)})}^{(s-wave)}={\Delta}_{CDW(\mathbf{Q}_{x(y)})}^{(1)}+{\Delta}_{CDW(\mathbf{Q}_{x(y)})}^{(2)}.\vspace{0.1cm}
\end{array}%
\right.$ 
\end{center}

\noindent As a result, the two-loop RG flow equations for the response functions associated with a potential instability of
the normal state towards a given ordered (i.e. symmetry-broken) phase then finally read

\vspace{-0.3cm}

\begin{align}
\Lambda \frac{d}{d\Lambda}\Delta_{PDW(\mathbf{Q}_{x(y)})}&=-\lambda(g_{3t}+g_{3v})\Delta_{PDW(\mathbf{Q}_{x(y)})}\nonumber\\
&+\eta\,\Delta_{PDW(\mathbf{Q}_{x(y)})},\label{PDW}\\
\Lambda \frac{d}{d\Lambda}\Delta_{CDW(\mathbf{Q}_{x(y)})}^{d-wave}&=\lambda(-2g_{3v}+g_{3t})
\Delta_{CDW(\mathbf{Q}_{x(y)})}^{d-wave}\nonumber\\
&+\eta\,\Delta_{CDW(\mathbf{Q}_{x(y)})}^{d-wave},\label{CDW_1}\\
\Lambda \frac{d}{d\Lambda}\Delta_{CDW(\mathbf{Q}_{x(y)})}^{s-wave}&=\lambda(2g_{3v}-g_{3t})
\Delta_{CDW(\mathbf{Q}_{x(y)})}^{s-wave}\nonumber\\
&+\eta\,\Delta_{CDW(\mathbf{Q}_{x(y)})}^{s-wave}.\label{CDW_2}
\end{align} 

\subsection{One loop RG}

To get a qualitative idea of the results in the RG analysis,
it can be very useful here to discuss first the results of the present model in a one-loop approximation,
before moving on to the full two-loop RG case. For this reason, we shall perform this now.
At one-loop RG order, we can calculate the PDW and CDW response functions close to the critical scale $\Lambda_c$
by simply substituting the ansatz $g_i (l)=C_i/(l_c-l)$ for the couplings
into Eqs. (\ref{PDW})-(\ref{CDW_2}) and setting the two-loop anomalous dimension contribution to zero
(i.e., $\eta=0$). By doing this, we immediately find two different regimes, which crucially depend on the initial value of the couplings for Hubbard-like interactions.
To understand this, we must compare two different energy scales which emerge in the present problem: the first one
is the critical energy scale $\Lambda_c$ set by the couplings that diverge in the low-energy limit and the other scale is the 
low-energy cutoff of the particle-hole and particle-particle bubbles (say, at the wavevector $\mathbf{q}=\mathbf{Q}_{x}$) defined by Eqs. (\ref{PH}) and (\ref{PP}), i.e., $v_F^R k_c(1-\tan\theta_R)/\lambda$. For simplicity, we will consider here the renormalized Fermi surface that emerges at two-loop RG order to make a qualitative estimate as to which of the above scales win over the other in the two physical regimes mentioned above.

\begin{figure}[t]
 \includegraphics[height=2.1in]{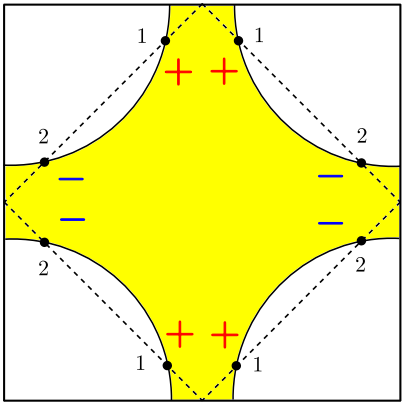}
 \caption{(Color online) The signs of the order parameters at all eight hot spots for a $d$-wave form factor. Instead, for an $s$-wave form factor,
 all signs become positive.}
 \label{fig:response_function}
\end{figure}

In the first regime, which occurs for weaker couplings in the present model (i.e. $g_i< g_c$, where $g_c$ is a critical coupling parameter), it is easy to verify that $v_F^R k_c > \Lambda_c$. This means that
the low-energy cutoff in the particle-hole and particle-particle bubbles at the incommensurate wavevector $\mathbf{q}=\mathbf{Q}_{x}$ will prevent those polarization functions in the corresponding susceptibilities
to diverge in the scaling limit. In this way, both PDW and CDW orders will remain subleading with respect to BDW and SSC, since these latter two orders clearly do not possess an infrared cutoff in their respective bubbles [see Eqs. (\ref{PH1}) and (\ref{PP1})] for the linearized dispersion assumed in the present model.
This result is clearly in agreement with several results obtained using different approaches concerning the spin-fermion model \cite{Chowdhury,Pepin_2}
and it has also been obtained by two of us within the context of the present fermionic hot-spot model in a previous work \cite{Freire5} and, subsequently, confirmed by other researchers in Ref. \cite{Seth}.

By contrast, in the second regime, and for our present purposes the most interesting one, the behavior of the system changes dramatically. This occurs for moderate couplings in the present model (i.e., $g_i> g_c$), since in this case it turns out that $\Lambda_c>v_F^R k_c$.
Most response functions calculated in this work will exhibit a divergence within such a ``high-temperature'' regime, which is due to the divergence of the couplings at one-loop RG order explained in the previous section. Consequently,
the PDW and CDW response vertices will now behave near the energy scale $\Lambda_c$ as a power-law
given by $\Delta_m(\Lambda)\sim (\Lambda-\Lambda_c)^{\alpha_m}$, with the exponents given approximately by 

\vspace{-0.4cm}

\begin{align}
 &\alpha_{PDW(\mathbf{Q}_{x(y)})}=-\lambda(C_{3t}+C_{3v})\approx -0.37,\\
 &\alpha^{d-wave}_{CDW(\mathbf{Q}_{x(y)})}=\lambda(-2C_{3v}+C_{3t})\approx -0.38,\\
 &\alpha^{s-wave}_{CDW(\mathbf{Q}_{x(y)})}=\lambda(2C_{3v}-C_{3t})\approx 0.38.
\end{align}

\noindent To establish a precise comparison among the different possibilities of ordering tendencies in the present model, 
we quote the previous one-loop results obtained by two of us in a previous work\cite{Freire5} regarding the critical exponents
of AF, BDW and SSC, which are given by

\vspace{-0.3cm}

\begin{align}
 &\alpha_{AF}=-\frac{1}{2}(C_{2}+C_{2x}+C_{3}+C_{3x}+2C_{3v})\approx -0.75,\\
 &{\alpha}^{d-wave}_{BDW}=\frac{1}{2}(2C_{1c}-C_{2c}+C_{3p}-2C_{3x})\approx -0.16,\\
 &\alpha^{d-wave}_{SSC}=\frac{1}{2}(C_{2c}+C_{1c}-C_{1x}-C_{2x})\approx -0.19.
\end{align}

\noindent We note from the above one-loop RG results that the exponents associated with CDW and PDW response functions
(both with a $d$-wave form factor) at the experimentally observed incommensurate wavevectors $\mathbf{Q}_{x(y)}$ turn out to be negative,
which implies that the two quantities diverge close to the
critical scale $\Lambda_c$ at this order. The only exception is the CDW order at wavevectors $\mathbf{Q}_{x(y)}$ with an $s$-wave form factor
whose exponent is positive and therefore the corresponding response vertex instead renormalizes to zero near $\Lambda_c$.
This latter order parameter turns out to be irrelevant in this case and, for this reason, we shall neglect
this order from here on.
In addition to this, we find, quite surprisingly, that the exponents associated with $d$-wave CDW and PDW at $\mathbf{Q}_{x(y)}$ 
turn out to be larger than the exponents related to both BDW and SSC orders
within this regime. We mention here that this qualitative trend will also hold at two-loop RG level.
Therefore, we establish here that the PDW and the experimentally relevant CDW indeed emerge 
as leading ordering tendencies compared to BDW/SSC in the present fermionic hot-spot model. 
With further hole doping, the absolute value of the wave vectors $\mathbf{Q}_{x(y)}$ that determine the modulation of the PDW and CDW orders will naturally decrease.
This dependence of the observed CDW wave vector on doping has some support from the experimental situation \cite{LeTacon}
and is suggestive that the Fermi surface is indeed playing some role in non-Lanthanum-based compounds.
Another very interesting feature coming out of the above one-loop RG result is the reasonable proximity of
the critical exponents associated with PDW and CDW.
As will become clear shortly, this is not a coincidence and in fact is a precursor of the appearance of a new emergent symmetry at low energies, which relates these two orders in
the present model. Finally, we point out that the above divergences of the response functions turn out to be an artifact of 
the one-loop approximation and, as we will see
next, two-loop order
terms will eliminate such singularities altogether. For this reason,
we now move on to the two-loop RG case that we are, of course, mostly interested in.

\subsection{Two loop RG}
 
As we have shown earlier in this paper, the renormalized couplings at two loops approach asymptotically only one IR-stable nontrivial 
fixed point (the HSFP), which controls the universal physics of the model at low energies. 
By integrating Eqs. (\ref{PDW})-(\ref{CDW_1}) in the vicinity of this fixed point, we obtain
that the calculated response vertices must obey power-laws described by $\Delta_{m}(\Lambda)\sim\Lambda^{\nu_{m}^{*}}$,
where the two-loop critical exponents are given by

\vspace{-0.4cm}

\begin{align}
 &\nu_{PDW(\mathbf{Q}_{x(y)})}=-\lambda(g^{*}_{3t}+g^{*}_{3v})+\eta^{*}\approx 3.22, \label{nu_1}\\
 &\nu^{d-wave}_{CDW(\mathbf{Q}_{x(y)})}=\lambda(-2g^{*}_{3v}+g^{*}_{3t})+\eta^{*}\approx 3.22, \label{nu_2}\\
 &\nu_{AF}=-\frac{1}{2}(g^{*}_{2}+g^{*}_{2x}+g^{*}_{3}+g^{*}_{3x}+2g^{*}_{3v})+\eta^{*}\approx 0.04,\label{nu_3}\\
 &{\nu}^{d-wave}_{BDW}=\frac{1}{2}(2g^{*}_{1c}-g^{*}_{2c}+g^{*}_{3p}-2g^{*}_{3x})+\eta^{*}\approx 0.30,\label{nu_4}\\
 &\nu^{d-wave}_{SSC}=\frac{1}{2}(g^{*}_{2c}+g^{*}_{1c}-g^{*}_{1x}-g^{*}_{2x})+\eta^{*}\approx 0.30.\label{nu_5}
\end{align}

\noindent From the above result, we conclude that at two-loop RG level the critical exponents of all response functions calculated in this work become
positive and, as a consequence, instead of a divergence displayed at one-loop order,
they now clearly scale down to zero at two loops. Physically speaking, this happens because of the two-loop self-energy feedback
(i.e., the pseudogap instability) onto the
RG equations that acts as a ``preemptive order'' that self-consistently
gaps all the correlations out, thus finally generating only short-range orders in the present model.
The two-loop-order corrections also implies that the quasiparticle weight $Z_{\psi}$ tends to be nullified at the hot spots in the present model and
that both uniform spin and charge susceptibilities tend to become suppressed in the low-energy limit, which was
previously obtained in Ref. \cite{Freire2}. This may either point to a partial truncation of the Fermi surface at the
hot spots (e.g., Fermi arcs) or it might also lead to a full reconstruction of the Fermi surface
into pockets. Thus, the properties of the critical theory obtained here at two loops capture some aspects of
the phenomenology exhibited by the underdoped cuprates at high temperatures.
Despite this, we note from Eqs. (\ref{nu_1})-(\ref{nu_5}) that the dominant antiferromagnetic spin fluctuations 
exhibit a large but finite correlation length $\xi_{AF}$.
This latter fact, however, disagrees with the experimental situation, where the correlation length of AF spin fluctuations
is observed to be only of the order of a few lattice constants in underdoped cuprates.

Another very interesting property that we obtain straightforwardly at two-loop RG level concerns one additional emergent $SU(2)$ pseudospin symmetry that
shows up naturally in the present model. 
Here we use the word emergent in the sense that the original fermionic hot spot model that we analyze in the present work is not invariant under
this given symmetry, but only at the novel two-loop fixed point we obtain that this symmetry appears. The first emergent $SU(2)$ symmetry
that relates the
$d$-wave SSC order and the $d$-wave BDW order parameter at the incommensurate wavevectors $\mathbf{\widetilde{Q}}=(\pm Q_0,\pm Q_0)$ was 
previously emphasized in
the context of the fermionic hot-spot model in Ref. \cite{Freire5}
and this is clearly evidenced by the equality of those respective critical exponents. This symmetry is descendant of the $SU(2)$ pseudospin
symmetry demonstrated by Metlitski and Sachdev \cite{Sachdev} and explored further by Efetov \emph{et al.} \cite{Efetov} for the case of the spin-fermion model
with a linearized dispersion.
In addition to this fact, we find that the critical exponents associated to PDW and CDW order at
the experimentally observed wavevectors $\mathbf{Q}_{x(y)}$ 
(both with a $d$-wave form factor) also turn out to be numerically equal to each other. The underlying
reason for this degeneracy is that these latter order parameters 
also turn out to be linked by a similar $SU(2)$ pseudospin symmetry
associated with particle-hole transformations in the present model (one for each pair of hot spots connected by $(\pi,\pi)$)
that maps the $d$-wave charge order at $\mathbf{Q}_{x(y)}$ onto the PDW order parameter at the same wavevector.
In the context of the spin-fermion model, this has been noted
by Pépin \emph{et al.} \cite{Pepin_2} (see also Ref. \cite{Pepin_3}) by including a small breaking of the tetragonal
symmetry (orthorhombicity) at a mean-field calculation and also, quite recently, by Wang \emph{et al.} \cite{Wang_2,Wang_3} 
using a Ginzburg-Landau theory,
in which case the CDW and PDW order parameters become components
of an $SO(4)$ PDW/CDW ``super-vector''.
Therefore, we reaffirm here this result also in the present context of the fermionic hot-spot model (but with no
breaking of $C_4$ lattice rotational symmetry)
by taking into account important higher-order fluctuation effects. This agrees with the observation
that the two emergent $SU(2)$ pseudospin symmetries established here 
manifest themselves in the present fermionic hot spot model coexisting with -- but not simply mediated by (see, e.g, Ref. \cite{Norman}) -- strong short-range AF spin fluctuations. Moreover, in an analogous way to the previous section, such an entangled PDW/CDW order also turns out to be subleading to BDW/SSC for weak couplings in the present model at two-loop RG level. However, as we increase the couplings to somewhat moderate interactions, we do find a tendency for PDW/CDW to become leading compared to BDW/SSC near the HSFP obtained in the present work. This last result would in principle agree with the experimental situation, in which the leading charge order that emerge in non-Lanthanum-based cuprates has indeed a modulation determined by $\mathbf{Q}_{x(y)}$.Therefore, it seems reasonable to speculate here that this new emergent $SU(2)$ symmetry relating PDW and CDW may indeed have profound consequences for the physics of the underdoped cuprates, as has been recently alluded to by many authors in the literature \cite{PALee,Wang_3,Fradkin,Agterberg}.

\section{Conclusions}

In the present work, we have investigated within a complete two-loop RG framework the fermionic hot spot model relevant to the phenomenology of the
cuprates, which describes excitations with a linearized dispersion
in the vicinity of eight hot spots (i.e., the points in momentum space in which 
the AF zone boundary intersects a putative underlying Fermi surface of the cuprate superconductors at low hole doping).
The present model can be seen as descendant of the Abanov-Chubukov spin-fermion model that, most importantly, includes
here all relevant interactions between the fermions and, for this reason, allows one to investigate on equal footing all of its possible instabilities 
from a weak to moderate coupling regime.

Here, we have explicitly studied the role played by two types of order in the model (CDW and PDW) in the
vicinity of the IR-stable nontrivial fixed point obtained at two loops. By analyzing
the CDW response at the experimentally relevant wavevectors $(\pm Q_0,0)$ and $(0,\pm Q_0)$, we were able
to establish that this charge order is short-ranged and has a predominant $d$-wave form factor, consistent with recent STM experiments \cite{Fujita}
and resonant x-ray scattering \cite{Comin}. We have also focused our attention on the so-called PDW order, 
which was recently proposed in the literature as
a potential candidate for the ``hidden'' order to describe the pseudogap phase observed in underdoped cuprates \cite{PALee,Agterberg,Fradkin,Wang_2},
since it may lead to a secondary order parameter that breaks
both time-reversal and parity symmetries.
In this respect, we have confirmed that the PDW order with the same modulation given by $\mathbf{Q}_{x(y)}$ emerges as an $SU(2)$-degenerate counterpart of
the CDW, which bears some resemblances with the results in the spin-fermion model obtained
by Pépin \emph{et al.} \cite{Pepin_2} at a mean-field calculation and also by Wang \emph{et al.} \cite{Wang_2}.
In such a case, the PDW and CDW order parameters should become components
of an SO(4) ``super-vector'' (hence the denomination PDW/CDW proposed in Refs. \cite{Pepin_2,Wang_2}).

For weaker couplings in the model, the PDW/CDW composite order (with a predominant $d$-wave form factor) is always subleading compared
to BDW/SSC. This is in agreement with several works in the literature that either consider a spin-fermion model with weak interactions \cite{Chowdhury,Pepin_2} or
start from the full lattice problem and analyze it within a mean-field approximation \cite{LaPlaca,Norman}. 
By contrast, as we increase the coupling of the model towards moderate values, the entangled PDW/CDW order
becomes leading compared to BDW/SSC, which clearly agrees with the experimental situation.
We point out that our present result resonates with a recent proposal by P. A. Lee \cite{PALee} that the CDW order experimentally observed in the underdoped cuprates could be interpreted in terms of a PDW, which could arise as a potential strong coupling instability in a theory of high-$T_c$ superconductivity from the perspective of doping a Mott insulator. Therefore, our present conclusion may also be seen in qualitative agreement with this point of view.

Finally, we point out that it is crucial to implement the RG method for the fermionic hot model 
at least at two-loop order or beyond, since we have 
shown here that the first nonanalytic contribution to the self-energy 
(which is responsible for many effects such as, e.g., the renormalization of the
single-particle renormalized Green's function and related quantities\cite{Freire5,Ferraz}, the renormalization of the FS, and others) 
only emerges at two loops.  
Physically speaking, this also happens because the two-loop self-energy feedback
(i.e., the pseudogap instability) onto the
RG equations acts to self-consistently
gap all the correlations out, thus eventually generating only short-range orders in the present model.
The two-loop-order corrections in addition imply that, for moderate interactions,
the quasiparticle weight tends to be nullified at the hot spots in the present model and
that both uniform spin and charge susceptibilities tend to become suppressed in the low-energy limit, which was
previously obtained in Ref. \cite{Freire2}. This may either point to a partial truncation of the Fermi surface at the
hot spots (e.g., Fermi arcs) or it might also lead to a full reconstruction of the Fermi surface
into pockets. Thus, the properties of the critical theory obtained here might be relevant to describe some aspects of
the phenomenology exhibited by the underdoped cuprates at high temperatures.
Further work along the lines of the one presented here would include to consider more realistic models in which we
couple, for instance, all the hot spots to
both the so-called ``lukewarm'' spots (i.e., the points that are reasonably close to the hot spots) and also to the cold spots in the nodal direction
to analyze how robust are the present results with respect to adding more degrees of freedom in the system.
Also, another direction that clearly deserves a future investigation will be
to calculate transport properties in the present model, such as electrical resistivity and
thermal conductivity. Since the quasiparticles are not well-defined in the present model,
one must use instead the so-called memory matrix formalism \cite{Forster} that, most importantly, does not assume the existence of quasiparticles at low energies.
This calculation of non-equilibrium properties has been initiated in recent years to discuss many 
strongly correlated systems such as, e.g.,
the theory of an Ising-nematic transition out of a metallic state which
breaks the lattice rotation symmetry but preserves translational symmetry \cite{Hartnoll}, a
spin-liquid model with a spinon Fermi surface coupled to a $U(1)$ gauge
field \cite{Freire6}, and lastly the theory of spin-density-wave quantum critical metals \cite{Hartnoll_2,Patel}. For this reason, we also plan to 
perform such an important investigation for the present fermionic hot-spot model in a future publication.

\begin{acknowledgments}
We want to thank E. Corrêa and A. Ferraz for useful discussions.
H.F. acknowledges financial support from FAPEG under grant No. 201210267001167 for this project.
V.S. de C. acknowledges financial support
from CAPES.
\end{acknowledgments}

\appendix

\section{}

In this appendix, we separate, for simplicity, the RG flow equations for the couplings $g_{1s}$, $g_{1r}$, $g_{1b}$, and $g_{2f}$ that
we display here up to one-loop order taking into account the effect of the renormalization of the FS. They are given by

\vspace{-0.3cm}

\begin{align}
&\Lambda\frac{d g_{1s}}{d\Lambda}=(g_{1c}+g_{2c})g_{1s}+(g_{1x}+g_{2x})g_{1r},\\
&\Lambda\frac{d g_{1r}}{d\Lambda}=(g_{1c}+g_{2c})g_{1r}+(g_{1x}+g_{2x})g_{1s},\\
&\Lambda\frac{d g_{1b}}{d\Lambda}=\lambda g_{1b}^2,\\
&\Lambda\frac{d}{d\Lambda}(2g_{2f}-g_{1b})=0.
\end{align}

\noindent As for the complete RG flow equations up to two loops (also taking into account the effect of the renormalization of the FS) 
of the other couplings in the present fermionic hot spot model described by the function $\beta_i=\Lambda d g_i^R/d\Lambda$,
they are given by the following expressions

\begin{figure*}[t]
 \includegraphics[height=3.3in]{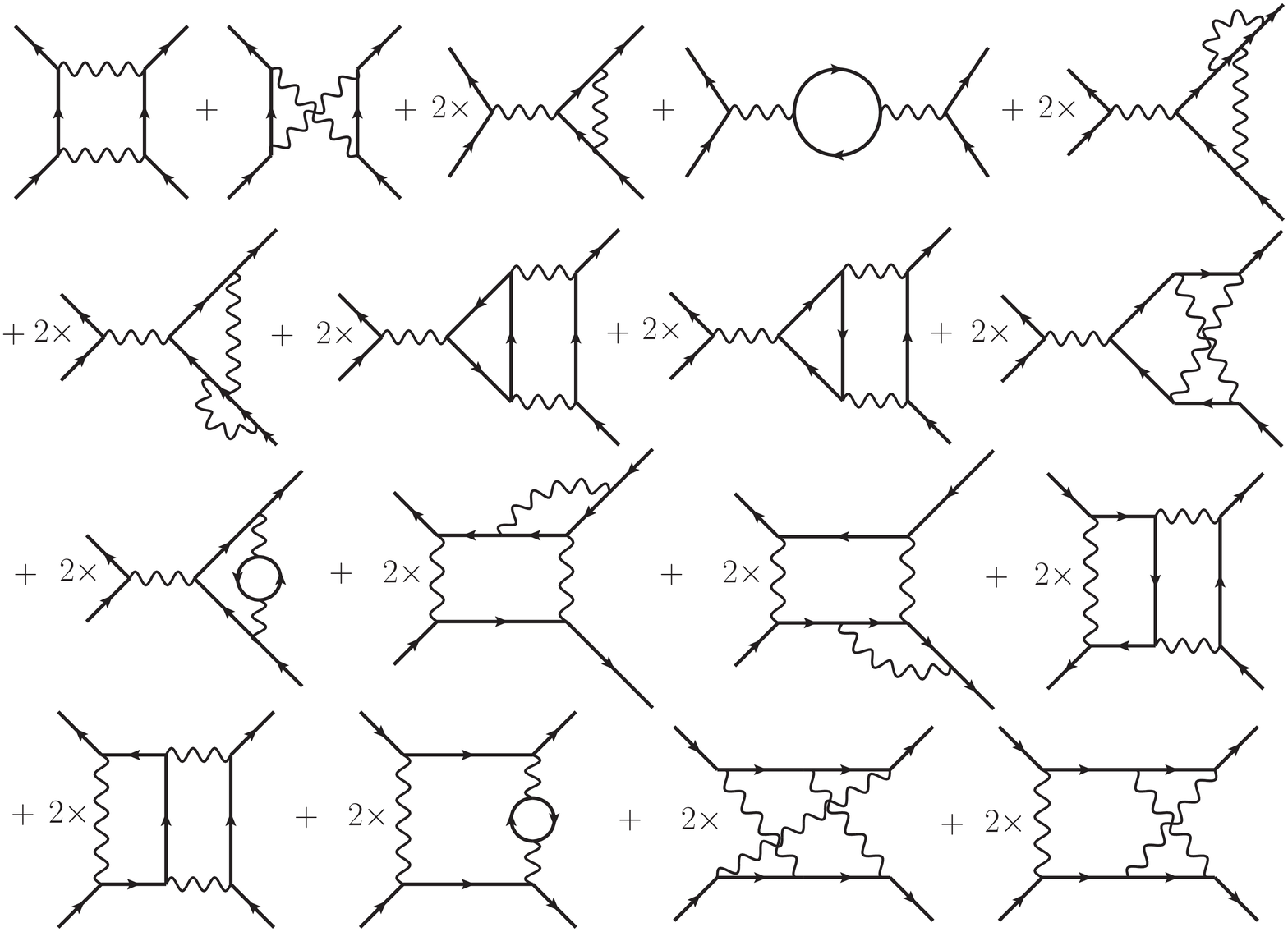}
 \caption{Relevant Feynman diagrams for the vertex corrections up
to two-loop order in the present model. Solid lines denote noninteracting fermionic single-particle Green's function, 
while the wavy lines correspond to the renormalized coupling constants.}
 \label{fig:Vertex_corrections}
\end{figure*}

\begin{widetext}

\begin{eqnarray}
\beta_{1}&=&g^{2}_{1}+g^{2}_{1x}+4g^{2}_{3t}+g^{2}_{3p}-g_{1x}g_{2x}-g_{3p}g_{3x}-4g_{3v}g_{3t}+\frac{1}{2}(g_{1x}g_{2x}-g^{2}_{2x}-g_{3p}g_{3x})g_{1c}\nonumber\\
&+&\frac{1}{2}(g^{2}_{1c}+g^{2}_{1}+g^{2}_{1x}+g^{2}_{2x}-g_{1x}g_{2x}-g_{3p}g_{3x}+g^{2}_{3p}+g^{2}_{3x})g_{1},\\
\beta_{2}&=&\frac{1}{2}\left(g^{2}_{1}-g^{2}_{2x}-g^{2}_{3}-g^{2}_{3x}\right)-2g^{2}_{3v}+\frac{1}{4}\left(g^{3}_{1}+g_{1c}g^{2}_{1x}+g_{1}g^{2}_{1c}\right)+\frac{1}{4}\left(2g_{2}-g_{1}\right)g^{2}_{3}\nonumber\\
&+&\frac{1}{4}\left[\left(2g_{2c}-g_{1c}\right)\left(g^{2}_{3p}+g^{2}_{3x}\right)-2g_{2c}g_{3p}g_{3x}+2g_{2}\left(g^{2}_{3p}+g^{2}_{3x}-g_{3p}g_{3x}\right)\right]\nonumber\\
&+&\frac{1}{2}(g_{2}-g_{2c})(g_{1x}^{2}-g_{1x}g_{2 x})+\frac{1}{2}g_{2}g_{2x}^{2}+\frac{1}{2}g_{2c}g_{2x}^{2},\\
\beta_{3}&=&(g_{1}-2g_{2})g_{3}-g_{1x}(g_{3x}-2g_{3p})-g_{2x}(g_{3p}+g_{3x})+4(g_{3t}-g_{3v})g_{3t}-2g^{2}_{3v}+\frac{1}{4}[(g_{1}-2g_{2})^{2}\nonumber\\
&+&(g_{1c}-2g_{2c})^{2}+2g^{2}_{1x}+2g^{2}_{2x}-2g_{1x}g_{2x}-2g_{3p}g_{3x}+2g^{2}_{3p}+2g^{2}_{3x}+g^{2}_{3}]g_{3},\\
\beta_{3t}&=&(2g_{1}-g_{2}+g_{3}+2g_{1x}-g_{2x}+2g_{3p}-g_{3x})g_{3t}
-(g_{1}+g_{3}+g_{1x}+g_{3p})g_{3v}+2\eta g_{3t},\\
\beta_{3v}&=&-(g_{2}+g_{3}+g_{2x}+g_{3x})g_{3v}+2\eta g_{3v},\\
\beta_{1c}&=&g^{2}_{1c}+g_{1x}g_{2x}+g^{2}_{1s}+g^{2}_{1r}+g^{2}_{3x}-g_{3p}g_{3x}+\frac{1}{2}\left(g_{1x}g_{2x}-g^{2}_{2x}-g_{3p}g_{3x}\right)g_{1}\nonumber\\
&+&\frac{1}{2}\left(g^{2}_{1c}+g^{2}_{1}+g^{2}_{1x}+g^{2}_{2x}-g_{1x}g_{2x} +g^{2}_{3p}+g^{2}_{3x}-g_{3p}g_{3x}\right)g_{1c},\\
\beta_{2c}&=&\frac{1}{2}\left(g^{2}_{1c}+g^{2}_{1x}+g^{2}_{2x}+2g^{2}_{1s}+2g^{2}_{1r}-g^{2}_{3p}\right)+\frac{1}{4}\left(g_{1}g^{2}_{1x}+g_{1c}g^{2}_{1}+g^{3}_{1c}\right)\nonumber\\
&+&\frac{1}{2}(g_{2c}-g_{2})(g_{1x}^{2}+g^{2}_{2x}-g_{1x}g_{2x})+\frac{1}{4}\left(2g_{2c}-g_{1c}\right)g^{2}_{3}\nonumber\\
&+&\frac{1}{4}\left[\left(2g_{2}-g_{1}\right)\left(g^{2}_{3p}+g^{2}_{3x}\right)-2g_{2}g_{3p}g_{3x}+2g_{2c}\left(g^{2}_{3p}+g^{2}_{3x}-g_{3p}g_{3x}\right)\right],\\
\beta_{1x}&=&g_{1c}g_{2x}+g_{2c}g_{1x}+2g_{1s}g_{1r}+2g_{1x}g_{1}-g_{2x}g_{1}-g_{1x}g_{2}+(g_{3p}-g_{3x})g_{3}+ 4g^{2}_{3t}-4g_{3v}g_{3t}\nonumber\\
&+&\frac{1}{2}\biggl(g_{1}g_{2c}+g_{1c}g_{2}-2g_{2c}g_{2}-\frac{g^{2}_{3p}}{2}-\frac{g^{2}_{3x}}{2}\biggr)g_{1x}+2\eta g_{1x},
\end{eqnarray}

\begin{eqnarray}
\beta_{2x}&=&g_{1c}g_{1x}+g_{2c}g_{2x}+2g_{1s}g_{1r}-g_{2}g_{2x}-g_{3}g_{3x}-g^{2}_{3v}+\frac{1}{2}\biggl(g_{1c}g_{1}g_{1x}-2g_{1c}g_{1}g_{2x}+g_{1c}g_{2}g_{2x}\nonumber\\
&+&g_{1}g_{2c}g_{2x}-2g_{2c}g_{2}g_{2x}-g_{1x}g_{3p}g_{3x}+g_{2x}g_{3p}g_{3x}-\frac{1}{2}g_{2x}(g_{3x}^{2}+g_{3p}^{2})\biggr)+2\eta g_{2x},\\
\beta_{3p}&=&(2g_{1}-g_{2c})g_{3p}+g_{1x}g_{3}+4g^{2}_{3t}-g_{2}g_{3p}-g_{2x}g_{3}-g_{1}g_{3x}-4g_{3v}g_{3t}+\frac{1}{2}\biggl(2g_{2c}g_{2}g_{3p}\nonumber\\
&+&g^{2}_{2x}g_{3x}-g_{1}g_{2c}g_{3p}-g_{1c}g_{2}g_{3p}-g_{1x}g_{2x}g_{3x}-g_{1c}g_{1}g_{3x}-\frac{g^{2}_{1x}g_{3p}}{2}\biggr)+2\eta g_{3p},\\
\beta_{3x}&=&(2g_{1c}-g_{2c})g_{3x}-g_{1c}g_{3p}-g_{2x}g_{3}-g_{2}g_{3x}-g^{2}_{3v}+\frac{1}{2}\biggl(2g_{2c}g_{2}g_{3x}+g^{2}_{2x}g_{3p}-g_{1}g_{2c}g_{3x}\nonumber\\
&-&g_{1c}g_{2}g_{3x}-g_{1x}g_{2x}g_{3p}-g_{1c}g_{1}g_{3p}-\frac{g^{2}_{1x}g_{3x}}{2}\biggr)+2\eta g_{3x},
\end{eqnarray}

\end{widetext}

\noindent where we have omitted the superscript $R$ in the renormalized dimensionless couplings of the model to not clutter up our notation
and $\eta$ is the two-loop anomalous dimension contribution. In Fig. 6, we show the corresponding Feynman diagrams for this calculation up to two loops.


\begin{thebibliography}{999}

\bibitem{Julien} T. Wu, H. Mayaffre, S. Kramer, M. Horvatic, C. Berthier, W. N. Hardy,
R. Liang, D. A. Bonn, and M.-H. Julien, Nature \textbf{477}, 191 (2011).

\bibitem{Julien2} T. Wu, H. Mayaffre, S. Krämer, M. Horvatic, C. Berthier, P. L. Kuhns, 
A. P. Reyes, R. Liang, W. N. Hardy, D. A. Bonn, and M.-H. Julien, Nat. Commun.
\textbf{4}, 2113 (2013).

\bibitem{Ghiringhelli} G. Ghiringhelli, M. Le Tacon, M. Minola, S. Blanco-
Canosa, C. Mazzoli, N. B. Brookes, G. M. De Luca, A.
Frano, D. G. Hawthorn, F. He, T. Loew, M. Moretti Sala,
D. C. Peets, M. Salluzzo, E. Schierle, R. Sutarto, G. A.
Sawatzky, E. Weschke, B. Keimer, and L. Braicovich, Science
\textbf{337}, 821 (2012).

\bibitem{Achkar} A. J. Achkar, R. Sutarto, X. Mao, F. He, A. Frano, S. Blanco-Canosa, M. Le Tacon, G. Ghiringhelli, L. Braicovich, M. Minola, M. Moretti Sala, C. Mazzoli, R. Liang, D. A. Bonn, W. N. Hardy, B. Keimer, G. A. Sawatzky, and D. G. Hawthorn, Phys. Rev. Lett. \textbf{109}, 167001 (2012).

\bibitem{Chang2} J. Chang, E. Blackburn, A. T. Holmes, N. B. Christensen, J. Larsen, J. Mesot, R. Liang, D. A. Bonn, W. N. Hardy, A. Watenphul, 
M. v. Zimmermann, E. M. Forgan, and S. M. Hayden, Nat. Phys. \textbf{8}, 871 (2012).

\bibitem{LeBoeuf}  D. LeBoeuf, S. Krämer, W. N. Hardy, R. Liang, D. A. Bonn, and Cyril Proust,
Nature Physics \textbf{9}, 79 (2013).

\bibitem{Hoffman} J. E. Hoffman, E. W. Hudson, K. M. Lang, V. Madhavan, H. Eisaki, S. Uchida,
and J. C. Davis, Science \textbf{295}, 466 (2002).

\bibitem{Yazdani} M. Vershinin, S. Misra, S. Ono, Y. Abe, Y. Ando, and A. Yazdani, Science
\textbf{303}, 1995 (2004).

\bibitem{Comin} R. Comin, R. Sutarto, F. He, E. da Silva Neto, L. Chauviere, A. Frano, R. Liang, W. N. Hardy, D. Bonn, Y. Yoshida, H. Eisaki, J. E. Hoffman, B. Keimer, G. A. Sawatzky, and A. Damascelli, Nat. Mater. \textbf{14}, 796 (2015).

\bibitem{Fujita} K. Fujita, M. H. Hamidian, S. D. Edkins, C. K. Kim, Y. Kohsaka, M. Azuma, M. Takano, H. Takagi, H. Eisaki, S.-i. Uchida, A. Allais, M. J. Lawler, E.-A. Kim, S. Sachdev, and J. C. S. Davis, Proc. Natl. Acad. Sci. \textbf{111}, E3026 (2014).

\bibitem{Comin2} R. Comin, A. Frano, M. M. Yee, Y. Yoshida, H. Eisaki, E. Schierle, E. Weschke,
R. Sutarto, F. He, A. Soumyanarayanan, Y. He, M. Le Tacon, I. S. Elfimov, J. E. Hoffman, G. A. Sawatzky, B. Keimer, and A. Damascelli, Science \textbf{343}, 390 (2014).

\bibitem{SilvaNeto} E. H. da Silva Neto, P. Aynajian, A. Frano, R. Comin, E. Schierle, E. Weschke, A. Gyenis, J. Wen, J. Schneeloch, Z. Xu, S. Ono, G. Gu, M. Le Tacon, 
and A. Yazdani, Science \textbf{343}, 393 (2014).

\bibitem{Taillefer} N. Doiron-Leyraud, C. Proust, D. LeBoeuf, J. Levallois, J.-B. Bonnemaison, R. Liang, D. A. Bonn, W. N. Hardy, and L. Taillefer, Nature \textbf{447}, 565 (2007).

\bibitem{Sebastian} S. E. Sebastian, N. Harrison, and G. G. Lonzarich, Reports on Progress in
Physics \textbf{75}, 102501 (2012).

\bibitem{Chubukov} A. Abanov and A. V. Chubukov, Phys. Rev. Lett. \textbf{84}, 5608 (2000);
Ar. Abanov, A. V. Chubukov, and J. Schmalian, Adv. Phys. \textbf{52}, 119 (2003).

\bibitem{Sachdev} M. A. Metlitski and S. Sachdev, Phys. Rev. B \textbf{82}, 075128 (2010).

\bibitem{Efetov} K. B. Efetov, H. Meier, and C. Pépin, Nature Physics \textbf{9}, 442 (2013).

\bibitem{LaPlaca} S. Sachdev and R. La Placa, Phys. Rev. Lett. \textbf{111}, 027202 (2013).

\bibitem{Efetov_2} H. Meier, C. Pepin, M. Einenkel, and K. B. Efetov, Phys. Rev. B \textbf{89}, 195115 (2014);
M. Einenkel, H. Meier, C. Pepin, and K. B. Efetov, Phys. Rev. B \textbf{90}, 054511 (2014).

\bibitem{Chowdhury} D. Chowdhury and S. Sachdev, Phys. Rev. B \textbf{90}, 134516 (2014).

\bibitem{Sau} J. D. Sau and S. Sachdev, Phys. Rev. B \textbf{89}, 075129 (2014).

\bibitem{Wang} Y. Wang and A. Chubukov, Phys. Rev. B \textbf{90}, 035149 (2014).

\bibitem{Allais} A. Allais, J. Bauer and S. Sachdev, Phys. Rev. B \textbf{90}, 155114 (2014).

\bibitem{Allais_2} A. Allais, J. Bauer and S. Sachdev, Indian Journal of Physics \textbf{88}, 905 (2014).

\bibitem{Tsvelik} A. M. Tsvelik and A. V. Chubukov, Phys. Rev. B \textbf{89}, 184515 (2014).

\bibitem{Kampf} S. Bulut, W. A. Atkinson, and A. P. Kampf, Phys. Rev. B \textbf{88}, 155132 (2013).

\bibitem{Chowdhury_2} D. Chowdhury and S. Sachdev, Phys. Rev. B \textbf{90}, 245136 (2014);  D. Chowdhury and S. Sachdev, arXiv:1501.00002.

\bibitem{Thomson} A. Thomson and S. Sachdev, Phys. Rev. B \textbf{91}, 115142 (2015).

\bibitem{Bulut} S. Bulut, A. P. Kampf, W. A. Atkinson, arXiv:1503.08896 (2015).

\bibitem{PALee} P. A. Lee, Phys. Rev. X \textbf{4}, 031017 (2014).

\bibitem{Agterberg} D. F. Agterberg, D. S. Melchert, and M. K. Kashyap,
Phys. Rev. B \textbf{91}, 054502 (2015).

\bibitem{Fradkin} E. Fradkin, S. A. Kivelson, and J. M. Tranquada,
Reviews of Modern Physics \textbf{87}, 457 (2015).

\bibitem{Wang_2} Y. Wang, D. F. Agterberg, and A. Chubukov, 
Phys. Rev. B \textbf{91}, 115103 (2015).

\bibitem{PALee_2} For a comprehensive review, see P. A. Lee, N. Nagaosa, and X. -G. Wen,
Rev. Mod. Phys. \textbf{78}, 17 (2006).

\bibitem{Bourges} B. Fauque, Y. Sidis, V. Hinkov, S. Pailhes, C. T. Lin, X. Chaud,
and P. Bourges, Phys. Rev. Lett. \textbf{96}, 197001 (2006).

\bibitem{Bourges_2} L. Mangin-Thro, Y. Sidis, A. Wildes, and P. Bourges, arXiv:1501.04919. 

\bibitem{Varma} C. M. Varma, Phys. Rev. B \textbf{73}, 155113 (2006).

\bibitem{Fulde} P. Fulde and R. A. Ferrell, Phys. Rev. \textbf{135}, A550 (1964).

\bibitem{Larkin} A. I. Larkin and Y. N. Ovchinnikov, Sov. Phys. JETP \textbf{20}, 762 (1965).

\bibitem{Greven} Y. Li, V. Baledent, N. Barisic, Y. Cho, B. Fauque, Y. Sidis, G. Yu, X. Zhao,
P. Bourges, and M. Greven, Nature \textbf{455}, 372 (2008).

\bibitem{Kapitulnik} J. Xia, E. Schemm, G. Deutscher, S. A. Kivelson, D. A. Bonn, W. N. Hardy, R. Liang, W. Siemons, G. Koster, M. M. Fejer,
and A. Kapitulnik, Phys. Rev. Lett. \textbf{100}, 127002 (2008).

\bibitem{Kapitulnik_2} H. Karapetyan, J. Xia, M. Hucker, G. D. Gu, J. M. Tranquada, M. M. Fejer, and A. Kapitulnik, Phys. Rev. Lett. \textbf{112},
047003 (2014).

\bibitem{Freire5} V. S. de Carvalho and H. Freire, Annals of Physics \textbf{348}, 32 (2014).

\bibitem{Seth} S. Whitsitt and S. Sachdev, Phys. Rev. B \textbf{90}, 104505 (2014).

\bibitem{Solyom} J. Sólyom, Adv. Phys. \textbf{28}, 201 (1979).

\bibitem{Peskin} M. E. Peskin and D. V. Schroeder, \emph{An Introduction to Quantum Field Theory} (Perseus Books, Cambridge, 1995).

\bibitem{Freire2} V. S. de Carvalho and H. Freire, Nuclear Physics B \textbf{875}, 738 (2013).

\bibitem{Ferraz} A. Ferraz, Phys. Rev. B \textbf{68}, 075115 (2003).

\bibitem{Freire} H. Freire, E. Correa, and A. Ferraz, Phys. Rev. B \textbf{71}, 165113 (2005).

\bibitem{Freire4} H. Freire, E. Correa, and A. Ferraz, Phys. Rev. B \textbf{78}, 125114 (2008).

\bibitem{LeTacon} S. Blanco-Canosa, A. Frano, E. Schierle, J. Porras, T.
Loew, M. Minola, M. Bluschke, E. Weschke, B. Keimer
and M. Le Tacon, Phys. Rev. B \textbf{90}, 054513 (2014).

\bibitem{Pepin_2} C. Pepin, V. S. de Carvalho, T. Kloss, and X. Montiel, Phys. Rev. B \textbf{90}, 195207 (2014).

\bibitem{Pepin_3} T. Kloss, X. Montiel, and C. Pepin, Phys. Rev. B \textbf{91}, 205124 (2015).

\bibitem{Norman} V. Mishra and M. R. Norman, arXiv:1502.02782v1.

\bibitem{Wang_3} Y. Wang, D. F. Agterberg, and A. Chubukov, Phys. Rev. Lett. \textbf{114}, 197001 (2015).

\bibitem{Forster} D. Forster, \emph{Hydrodynamic Fluctuations, Broken Symmetry, and Correlation Functions} (Benjamin, Cambridge, 1975).

\bibitem{Hartnoll} S.A. Hartnoll, R. Mahajan, M. Punk, and S. Sachdev, Phys. Rev. B \textbf{89} (2014) 155130.

\bibitem{Freire6} H. Freire, Annals of Physics \textbf{349}, 357 (2014).

\bibitem{Hartnoll_2} S. A. Hartnoll, D. M. Hofman, M. A. Metlitski and S. Sachdev, Phys. Rev. B
\textbf{84}, 125115 (2011).

\bibitem{Patel} A. A. Patel and S. Sachdev, Phys. Rev. B \textbf{90}, 165146 (2014).







\end{thebibliography}
\end{document}